\documentclass[journal]{IEEEtran}
\ifCLASSINFOpdf
   \usepackage[pdftex]{graphicx}
\else
  \usepackage[dvips]{graphicx}
\fi
\ifCLASSOPTIONcompsoc
  \usepackage[caption=false,font=normalsize,labelfont=sf,textfont=sf]{subfig}
\else
  \usepackage[caption=false,font=footnotesize]{subfig}
\fi
\usepackage{url}

\begin{document}
%
\title{The Effects of Radiation Damage on CLYC Performance }
%
%
%

\author{Katherine E. Mesick,
        Daniel D.S. Coupland,
        Suzanne F. Nowicki,
        and~Laura C. Stonehill,
\thanks{K.E. Mesick, D.D.S. Coupland, and S.F. Nowicki are with the Space Science and Applications group at Los Alamos National Laboratory, Los Alamos, NM 87545 USA. Corresponding author email: kmesick@lanl.gov}
\thanks{L.C. Stonehill is with the Intelligence \& Emerging Threats program office at Los Alamos National Laboratory, Los Alamos, NM 87545 USA.}
\thanks{This work was supported by Los Alamos National Laboratory Directed Research \& Development.  Los Alamos National Laboratory is operated by Los Alamos National Security, LCC under contract to the Department of Energy's National Nuclear Security Administration.}
\thanks{Manuscript received November 10, 2017.}}

%
%

\markboth{}%
{}
%



\maketitle

\begin{abstract}
Cs$_2$LiYCl$_6$:Ce$^{3+}$ (CLYC) is a new scintillator that is an attractive option for applications requiring the ability to detect both gamma rays and neutrons within a single volume.  It is therefore of interest in applications that require low size, weight, or power, such as space applications.  The radiation environment in space can over time damage the crystal structure of CLYC, leading to reduced performance.  We have exposed 2 CLYC samples to 800 MeV protons at the Los Alamos Neutron Science Center, one to approximately 10 kRad and one to approximately 100 kRad.  We measured the pulse shapes and amplitudes, energy resolution, and figure of merit for pulse-shape discrimination before and after irradiation.  We have also measured the activation products and monitored for room-temperature annealing of the irradiated samples.  The results of these measurements and the impact of radiation damage on CLYC performance is presented.
\end{abstract}


%
\IEEEpeerreviewmaketitle

\section{Introduction}
\begin{figure}[!ht]
\centering
\includegraphics[width=0.45\textwidth]{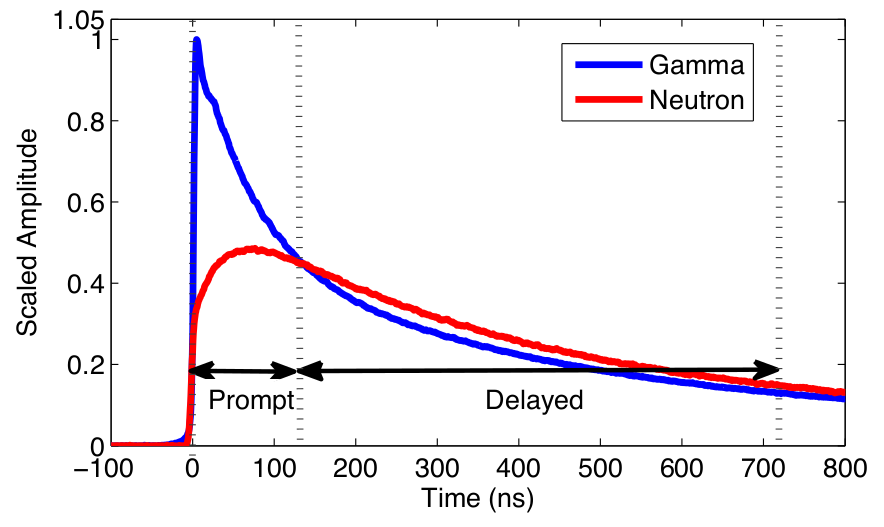}
\caption{CLYC pulse shapes in response to gamma and thermal neutron radiation.}
\label{fig:clyc_pulse}
\end{figure}

\begin{figure}[!ht]
\centering
\includegraphics[width=0.45\textwidth]{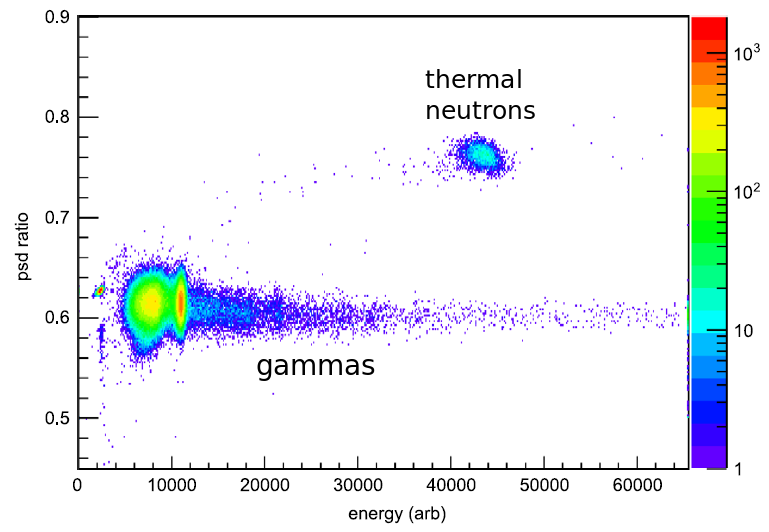}
\caption{Pulse-shape discrimination for CLYC, exhibiting distinct regions for gamma and thermal neutron events.}
\label{fig:clyc_psd}
\end{figure}

%
%
%
%
\IEEEPARstart{C}{} s$_2$LiYCl$_6$:Ce$^{3+}$ (CLYC) has gained much attention recently for its ability to detect both gamma rays and neutrons within a single volume.  It has excellent linearity and light yield, provides superior resolution to NaI and BGO, and provides thermal neutron sensitivity through neutron capture on $^6$Li with a Q-value of 4.8 MeV.  Different characteristic time structures for gamma- and neutron-induced pulses in CLYC, shown in Fig.~\ref{fig:clyc_pulse}, allow for pulse-shape discrimination (PSD) to be achieved, as shown in Fig.~\ref{fig:clyc_psd}, through integrating different regions of the pulse shapes.

CLYC scintillator is an attractive option for many applications.  In particular, for applications where size, weight, and power (SWaP) are important considerations, CLYC can provide significant improvement over current instruments.  One area which can benefit significantly from CLYC is space applications, however, CLYC does not currently have any space heritage.  It is thus important to demonstrate the survivability of CLYC in the space environment.  We have studied extensively the thermal dependence of CLYC coupled to photomultipliers~\cite{budden2013} and silicon photomultipliers~\cite{mesick2015}.  Other primary considerations for space-flight readiness are survivability of the launch environment and the space radiation environment.

\IEEEpubidadjcol

This work focuses on how the space radiation environment may impact the performance of CLYC scintillator.  The total accumulated dose in this environment, which can be 10 -- 100 kRad per year, may cause reduced performance that may be due to \cite{Zhu1998} the formation of color centers that reduce light output, activation and afterglow leading to increased noise, and/or damage to the scintillation mechanism.

\section{Methods}

\begin{figure}[!ht]
\centering
\includegraphics[width=0.45\textwidth]{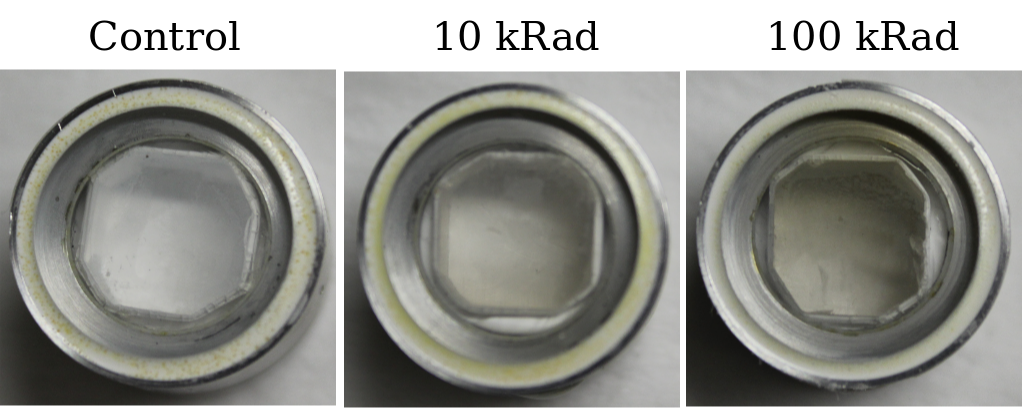}
\caption{CLYC samples showing discoloration in the irradiated crystals.}
\label{fig:samples}
\end{figure}

We irradiated two CLYC samples (1 cm cuboids) at the Los Alamos Neutron Science Center (LANSCE) facility with 800 MeV protons, one to approximately 10 kRad dose (4.8$\times$10$^{11}$ protons) and one to approximately 100 kRad dose (4.8$\times$10$^{12}$ protons).  A third identical CLYC sample was not irradiated as a control.  Discoloration of the irradiated CLYC samples can be seen in Fig.~\ref{fig:samples}.

The performance of the CLYC scintillator before and after the irradiation was quantified by measurements of the 662 keV photopeak resolution, the neutron and gamma waveform pulse shapes and amplitudes, and the ability to obtain a good figure of merit in the pulse-shape discrimination between neutron and gamma events.  The scintillators were read out with a Hamamatsu R11265-100 PMT.  The activation products were also measured 2 weeks after the irradiation with a high-purity germanium detector.

\section{Selected Results \& Discussion}

\begin{figure}[!h]
\centering
\includegraphics[width=0.45\textwidth]{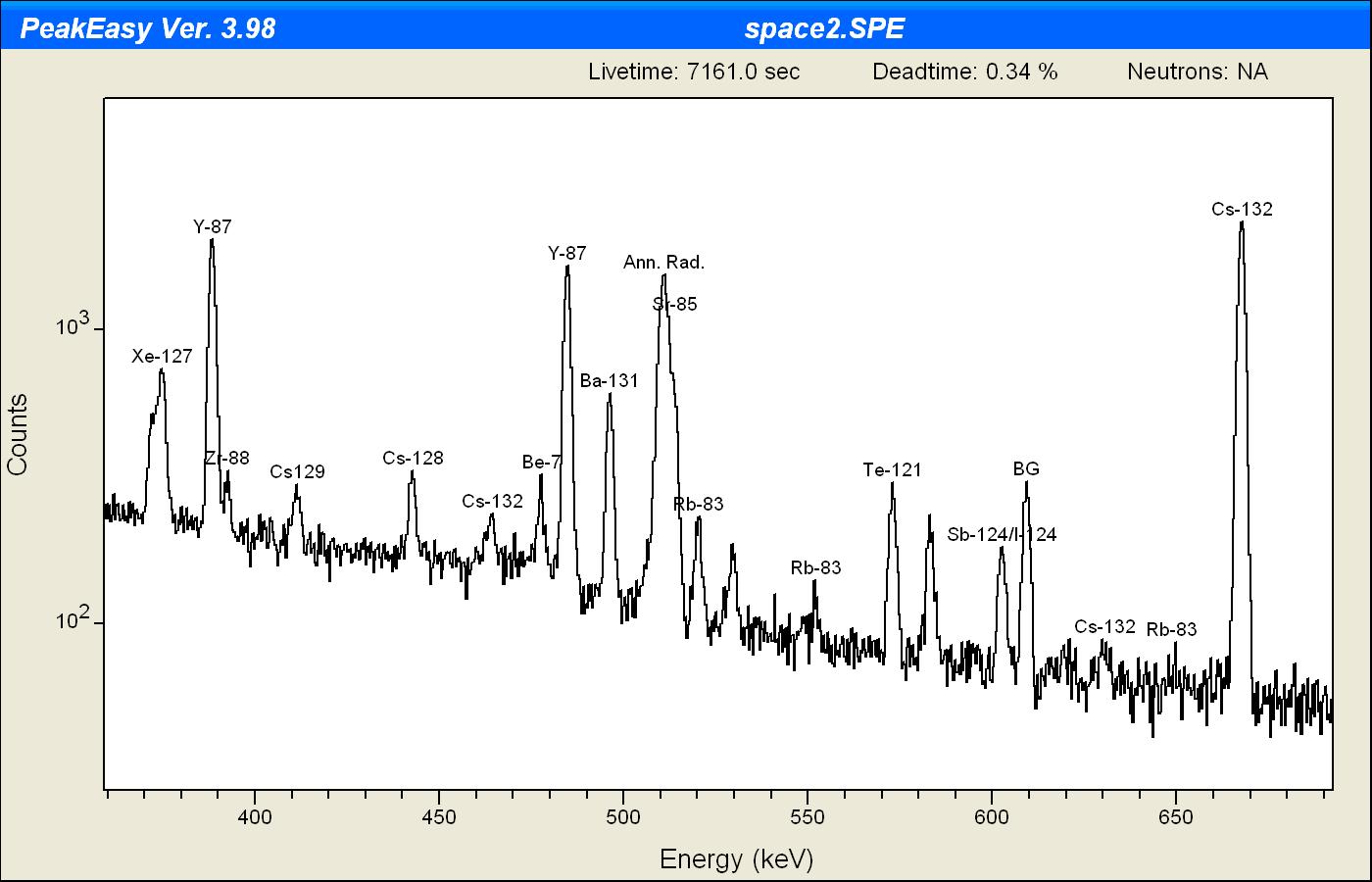}
\caption{Example of CLYC activation spectrum measured with HPGe after irradiation.}
\label{fig:activation}
\end{figure}

\begin{figure}[!ht]
\centering
\includegraphics[width=0.45\textwidth]{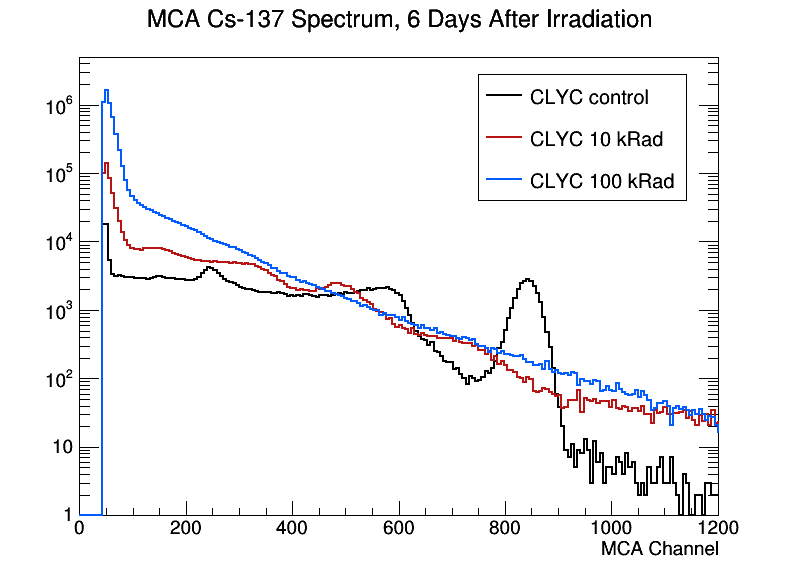}
\caption{$^{137}$Cs spectrum measured 6 days after irradiation showing large background from activation products.}
\label{fig:res_immediate}
\end{figure}

\begin{figure}[!ht]
\centering
\includegraphics[width=0.45\textwidth]{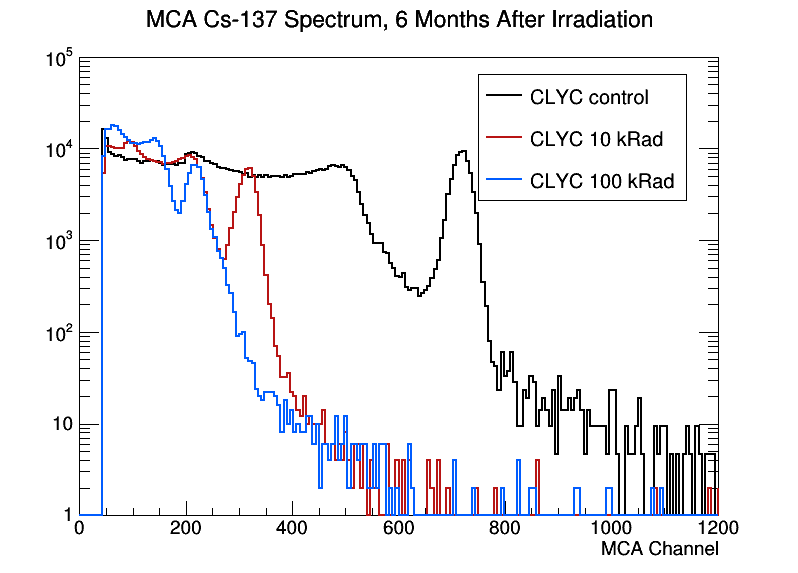}
\caption{$^{137}$Cs spectrum measured 6 months after irradiation showing reduced light output.}
\label{fig:res_6mos}
\end{figure}

The activation spectrum of CLYC after irradiation is shown in Fig.~\ref{fig:activation}. 
Most of the activation lines are attributable to the activation of the Cesium and Yttrium within CLYC or the housing material and have short half-lives from days to weeks.  This activation and a reduction in light output leads to noise and poor resolution of CLYC immediately after irradiation, as indicated in Fig.~\ref{fig:res_immediate}.  Here the MCA spectrum as measured in the presence of a $^{137}$Cs source shows the control with a 5.5\% full-width half-maximum photopeak resolution at 662 keV, while the irradiated samples show degraded performance due to the abundance of short-lived activation products.  After these activation products have decayed, a repeat measurement of $^{137}$Cs, shown in Fig.~\ref{fig:res_6mos}, shows the 662~keV photopeak visible for the irradiated CLYC samples, however with reduced light output and therefore worse resolution.

Pulse shapes were also digitized with an Agilent Acqiris DC282 waveform digitizer sampling at 2~GSamples/sec to assess the impact on pulse-shape discrimination.  Analysis of this data suggests that while overall decreased light output occurs, the pulse shapes are unchanged.  This results in little impact on the separation of the gamma- and neutron-induced events in the figure of merit.  However, the figure of merit does degrade in the irradiated CLYC crystals due to a broadening of the neutron PSD parameter.  

There was no evidence observed of room temperature annealing of the irradiated CLYC samples.  High-temperature annealing will be investigated in future measurements.

\ifCLASSOPTIONcaptionsoff
  \newpage
\fi

\end{document}